\documentclass[
reprint,
aps,
longbibliography
]{revtex4-1}
\usepackage[T1]{fontenc}
\usepackage[utf8]{inputenc}
\usepackage[english]{babel}
\usepackage{braket}
\usepackage{mathtools}	
\usepackage{amssymb}
\usepackage{amsmath}
\usepackage{dsfont}
\usepackage{graphicx}
\usepackage{hyperref}
\usepackage{siunitx}
\usepackage{xcolor}

\makeatletter
\renewcommand*{\fnum@figure}{{\normalfont\bfseries \figurename~\thefigure}}
\renewcommand*{\@caption@fignum@sep}{\textbf{. }}
\makeatother

\begin{document}
\title{Global optimization of quantum dynamics with AlphaZero deep exploration }
\author{Mogens Dalgaard}
\author{Felix Motzoi}
\author{Jens Jakob S{\o}rensen}
\author{Jacob Sherson}

\affiliation{Department of Physics and Astronomy, Aarhus University, Ny
	Munkegade 120, 8000 Aarhus C, Denmark}

\date{\today}
\begin{abstract}
While a large number of algorithms for optimizing quantum dynamics for different objectives have been developed, a common limitation is the reliance on good initial guesses, being either random or based on heuristics and intuitions. Here we implement a \emph{tabula rasa} deep quantum exploration version of the Deepmind AlphaZero algorithm for systematically averting this limitation. AlphaZero employs a deep neural network in conjunction with deep lookahead in a guided tree search, which allows for predictive hidden variable approximation of the quantum parameter landscape. To emphasize transferability, we apply and benchmark the algorithm on three classes of control problems using only a single common set of algorithmic hyperparameters. AlphaZero achieves substantial improvements in both the quality and quantity of good solution clusters compared to earlier methods. It is able to spontaneously learn unexpected hidden structure and global symmetry in the solutions, going beyond even human heuristics. 
\end{abstract}

\maketitle

Recent progress on technologies with quantum speedup focuses largely on optimizing dynamical quantum cost functionals via a set of external classical parameters. Such research includes quantum variational eigensolvers \cite{kandala2017hardware}, annealers \cite{johnson2011quantum}, simulators \cite{Lloyd1996, SimulationRMP2014}, circuit optimization \cite{Bocharov2015, Motzoi2017}, optimal control theory \cite{Warren1993Dream, khaneja2005optimal, GlaserCat2015}, and Boltzmann machines  \cite{biamonte2017quantum}. The minimized functional could be for example the energy of a simulated system, or the distance to a quantum computational gate. 

A shared algorithmic feature is domain knowledge about where to search, such as near the Hartree-Fock Ansatz for variational eigensolvers, or in the analytical gradient direction. An open question in optimization research is how much this specialized approach can be supplanted by a problem-agnostic methodology: One which does not require expert knowledge, avoiding both the overhead in human labour \cite{Sorensen2016} and the potential for local, suboptimal trapping \cite{pechen2011there,de2013closer,zhdanov2015role}. In other words, an autonomous machine learning approach has the potential to plan its solutions both strategically and tactically.

It has been argued that, due to the inherent smoothness of unitary quantum physics \cite{Hardy2001}, local \emph{exploitation} of quantum dynamics can be sufficient for efficiently finding good solutions \cite{Rabitz2004Traps}. Local search has been especially successful in the well-established field of Quantum Optimal Control Theory (QOCT), enjoying a half century of continued progress in NMR \cite{Freeman1987NMRbook}, quantum chemistry \cite{Warren1993Dream, Tannor2007ChemBook}, and spectroscopy \cite{Kawashima1995Spec}. This has culminated in Hessian extraction approaches \cite{quantumLBFGS} that generally outperform other local methods \cite{machnes2011comparing, sorensen2018quantum}.

Yet, similar to classical NP-complete problems \cite{Cheeseman1992NPtransition}, quantum functionals can suffer a phase transition \cite{bukov2018PhasesReinforcement} from easier to "needle in a haystack" instances that require global \emph{exploration} of parameters.
Mounting evidence has shown that critically constrained dynamics lead to such complexity \cite{Zahedinejad2014DEGA,Sorensen2016, Moore2012,bukov2018PhasesReinforcement}, especially as QOCT has veered into high-precision quantum computation \cite{Negrevergne2006nmr12}, circuit compilation \cite{Shende2006compil}, and architecture design \cite{Goerz2017fab}. It is therefore crucial to balance resources for exploitation of smooth, local quantum landscapes with state-of-the-art classical methods for domain-agnostic exploration.

In the literature, dynamics optimization is characterized by a lookahead-depth,
i.e.~how far into the future one plans current actions. A shallow depth may broaden exploration, a strategy typically found in Reinforcement Learning (RL) \cite{sutton2011reinforcement}. This has been powerfully combined with Deep Neural Networks (DNN) \cite{mnih2015human,lillicrap2015continuous,schulman2017proximal,salimans2017evolution,mania2018simple} and applied recently to quantum systems \cite{carleo2017solving, bukov2018reinforcementFloquet,zhang2018automatic,fosel2018reinforcement,niu2018universal, albarran2018measurement, an2019deep, xu2019transferable}. 
Unfortunately, single-step lookaheads are inherently local and thus require a slower learning rate, with no performance gain found over full-depth, domain-specialized (Hessian approximation) methods in QOCT. Other full-depth methods have also had mixed success, e.g. Genetic Algorithms \cite{GAchem1995, liebermann2016optimal} and Differential Evolution \cite{Zahedinejad2014DEGA}, but they typically require careful fine-tuning since they are based on \emph{ad-hoc} heuristics rather than being mathematically rooted.

A recent stunning breakthrough has been due to the \emph{AlphaZero} class of algorithms \cite{silver2016mastering,silver2017mastering,silver2017masteringchess}. AlphaZero has already effectively outclassed all adversaries in the games of Go, Chess, Shogu, and Starcraft. The key to the success of AlphaZero was the combination of a Monte Carlo tree search with a one-step lookahead DNN. As a result, the lookahead information from far down the tree dramatically increases the trained DNN precision, and together they compound to produce much more focused and heuristic-free exploration.

\begin{figure*}[htp]
	\centering
	\includegraphics[]{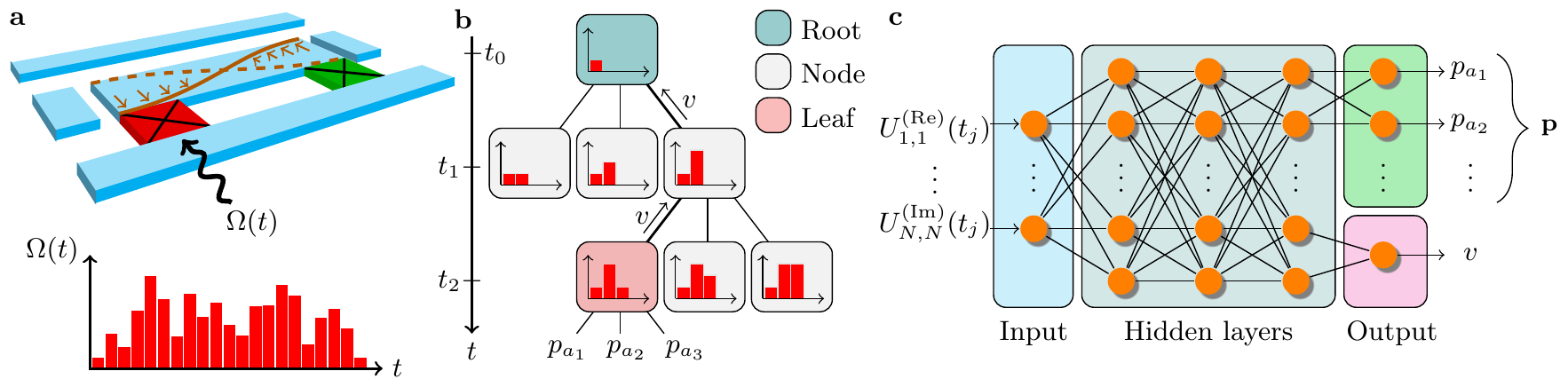}
	\caption{\textbf{a} Circuit-QED architecture consisting of two qubits (colored boxes) mounted on either side of a transmission line resonator. The first qubit is directly driven at the resonance frequency of the second one for a cross-resonance gate. An example of a piecewise constant pulse is depicted below the setup. \textbf{b} The schematics of a Monte Carlo tree search. Here the nodes are depicted as pulse sequences and the edges as lines. A single search consists of a forward propagation, expansion, and a back-pass (see text). \textbf{c} The neural network architecture used in AlphaZero. The network takes the state (unitary) as input and outputs probabilities for selecting individual actions $\mathbf{p} = \{ p_1, p_2, \ldots \}$ and an estimate of the final score (fidelity) $v$. }
	\label{fig:Transmon_treesearch_neuralnetwork}
\end{figure*}

Here, we implement and benchmark a QOCT version of AlphaZero for optimizing quantum dynamics. We characterize improvements in learning and exploration compared to traditional methods.  We find a crossover from difficult problems where AlphaZero learning alone is ideal and those where a combination of deep exploration and quantum-specialized smooth exploitation is optimal. We show this leads to  a dramatic increase in both the quality and quantity of good solution clusters. Our AlphaZero implementation retains the \emph{tabula rasa} character  of Ref. \cite{silver2017mastering} in two important respects. Firstly, it efficiently learns to solve three different optimization problem classes using the same algorithmic hyperparameters. Secondly, we demonstrate that AlphaZero is able to identify quantum-specific heuristics in the form of hidden symmetries without the need for expert knowledge.

\section{Results}
\subsection{Unified quantum exploration algorithm}

	\begin{figure*}[htp]
		\centering
		\includegraphics[]{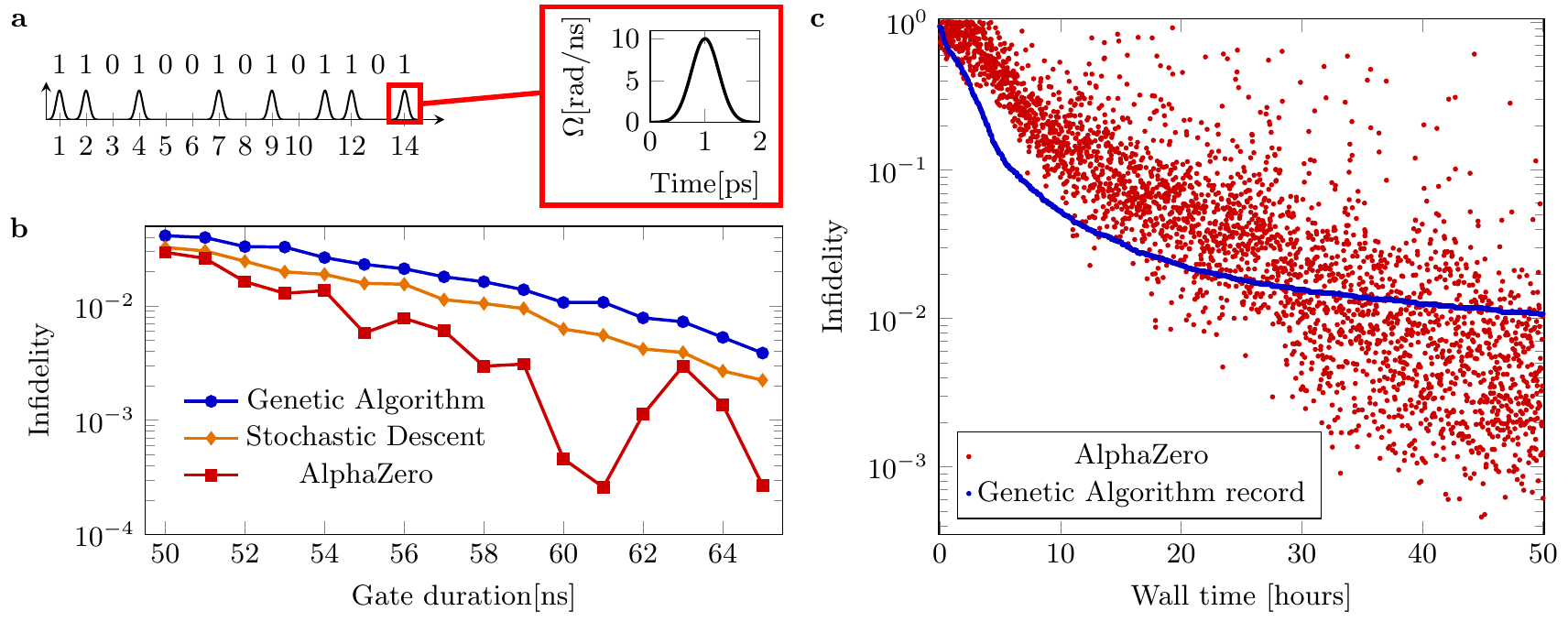}
		\caption{\textbf{a} Illustration of how a SFQ pulse train can be encoded into a bit string along with a zoom-in that depicts the exact shape of a single pulse. 
			\textbf{b} Infidelity ($1-\mathcal{F}$) as a function of gate duration for different discrete optimization algorithms. \textbf{c} A comparison between the infidelities obtained by AlphaZero and the GA at $60\si{\nano \second}$. For AlphaZero, each dot represents the infidelity obtained at the end of a unique episode, while for the GA each dot represents the highest scoring member in the population after each iteration.
		}
		\label{fig:SFQ}
	\end{figure*}

In this work, we seek to obtain pulse sequences that can unitarily steer a quantum system towards given desired dynamics. For our purposes, we quantify this task through the state-averaged overlap fidelity $\mathcal{F}(U(t))$ with respect to  a target unitary $\hat{U}_{\text{target}}$,
\begin{align} \label{eq:Fidelity}
\mathcal{F}(U(t)) = \bigg |
\frac{1}{\text{dim}} 
\text{Tr}
\big[
\hat{U}^{\dagger}(t) \hat{U}_{\text{target}} 
\big] 
\bigg |^2.
\end{align}
Here, $U(t)$ denotes the time evolution operator of the system, which solves the Schr\"odinger equation. We fix for concreteness our physical architeture as superconducting circuit QED \cite{ClarkSuperReview}, being both a highly tunable and potentially scalable architecture, with potential near-term applications  \cite{preskill2018quantum}. The system is chosen to be a resonator-coupled two-transmon system, as depicted in Fig.~\ref{fig:Transmon_treesearch_neuralnetwork}a. Here the transmon qubits are mounted on either side of a linear resonator and we drive the first qubit with an external control $\Omega$, which could be a piecewice constant pulse as depicted in the bottom of the figure. The system dynamics are governed by the Hamiltonian \cite{magesan2018effective}
\begin{align}
    \hat{H}(t) = \Delta \hat{b}_1^{\dagger} \hat{b}_1 + 
    J (\hat{b}_1^{\dagger} \hat{b}_2 + \hat{b}_1 \hat{b}_2^{\dagger})
    + \Omega(t) (\hat{b}_1^{\dagger} + \hat{b}_1),
    \label{eq:Hamiltonian}
\end{align}
where $b_j$ is the qubit-lowering operator for transmon $j$, and the external control $\Omega(t)$ is shaped by the optimization algorithm to maximize \eqref{eq:Fidelity}, with $\hat{U}_{\text{target}}=\sqrt{ZX}$ being a standard entangling gate.  $\hat{U}_{\text{target}}$ with single qubit gates form a universal gate set, e.g.~, for quantum computation on a surface code circuit-QED layout. We fix the parameters to be within typical experimental values (see e.g. Refs. \cite{leek2009using,sheldon2016procedure}) for the qubit-qubit coupling $J/2\pi = 5 \si{\mega \hertz}$ and the detuning $\Delta /2\pi = 0.35 \si{\giga \hertz}$.

We consider three optimization classes to test a unified AlphaZero algorithm and benchmark it against both domain-specialized and domain-agnostic algorithms. These three correspond to control parameters $\Omega(t)$ that are digital, i.e.~taken from a discrete set of possibilities; that can vary continuously as a function of continuous but highly-filtered controls; and lastly, piecewise constant controls, which is standard in the QOCT approximation.

Within the RL framework, an autonomous agent must interact with an environment that at each time step $t$ inhabits a state $s_t$. Here we choose the unitary $\hat{U}(t)$ to represent this state. The agent then alters the unitary at each time step $t$ by applying an action $a_t$ (here $\Omega(t)$) that transforms the unitary $\hat{U}(t) \rightarrow \hat{U}(t+\Delta t)$. The purpose of the agent is to maximize an expected score $z$ at final time $T$, which we choose to be the fidelity $z=\mathcal{F}(U(T))$. This is done by implementing a probabilistic policy $\boldsymbol{\pi}(s) = (\pi_{a_1}, \pi_{a_2},\ldots)$, which maps states $s$ to probabilities of applying actions, i.e.~$\pi_{a} = \text{Pr}(a|s)$. The agent attempts to improve the policy by gradually updating it with respect to the experience it gains.

Fig.~\ref{fig:Transmon_treesearch_neuralnetwork}b, c illustrate the tree search and the neural network for AlphaZero, respectively. The upper output of the neural network approximates the present policy for a given input state, i.e.~$p_a\sim\pi_a$. Meanwhile, the lower output provides a value function which estimates the expected final reward, that is $v(s_t)\sim \mathcal{F}(T)$. Both functions use only information about the current state and suffer from being lower-dimensional approximations of extremely high dimensional state and action spaces. The insight of the AlphaZero algorithm is to supplement the predictive power of the value function $v(s_t)$ with retrodictive information coming from future action decisions in a Monte Carlo search tree. The tree depicted in Fig.~\ref{fig:Transmon_treesearch_neuralnetwork}b consists of nodes, which represent states (here depicted as pulses) and edges, which are state-action pairs (depicted as lines). At each branch in the tree, the algorithm chooses actions based on a combination of those with the highest expected reward and the highest uncertainty, a measure of which edges remain unexplored. Whenever new states (called leaf-nodes) are explored, the neural network is used to estimate the value of that node, and the information is propagated backward in the tree to the root node. The forward and backward traversals of the tree are described in greater detail in Methods.

In the manner described above, the predictive nature of the network is able to inform choices in the tree while the retrodictive information coming back in time is able to give better estimates of the state values already explored, which are then used to train the network.  This reinforcing mechanism is thus able to globally learn about the parameter landscape by choosing the most promising branches while effectively culling the vast majority of the rest. The result is neither an exhaustive sampling at full depth, which would yield the true landscape albeit at a computationally untenable cost, nor is it an exhaustive sampling at shallow depth, which would require a prohibitively slow learning rate for information from the full depth of the tree to propagate back. Instead, AlphaZero intelligently balances the depth and the breadth of the search below each node. While the hidden-variable approximation given by the neural network and MC tree are certainly not exhaustive and cannot find solutions with exponentially small footprint, it is nonetheless able to discover patterns and learn an effective global policy strategy that produces robust, heterogeneous classes of promising solutions. In our implementation we restrict AlphaZero such that it can only find new unique solutions, which is done by cutting of branches in the tree that have previously been fully explored.  

In what follows we apply the algorithm with unified hyperparameters to three optimization classes: Discrete, continuous, and continous with strong constraints.  The three problem types accentuate different optimization strategies. In the discrete optimization case, we show how AlphaZero stands up against other domain-agnostic methods (where the gradient is not defined) and compare their abilities to learn structures in the parameters. For the constrained continuous pulses, we validate the hypothesis that the analytical gradient, while computable, is highly inefficient and indeed unable to find near global solutions that are at least as good as those found by AlphaZero. Finally, in the continuous-valued piecewise-constant case, we show the balance between state-of-the-art physics-specialized  and agnostic AlphaZero approaches. We show that the combination of exploration and exploitation is able to produce new clusters of high-quality solutions that are otherwise highly unlikely to be found, while learning hidden problem symmetry.

	\subsection{Digital gate sequences}

	As a first application with AlphaZero, we demonstrate optimal control using Single Flux Quantum (SFQ) pulses \cite{mcdermott2014accurate, liebermann2016optimal,  li2019scalable}.
	 The aim is to control the quantum system by using a pulse train that consists of individual, very short pulses typically in the pico-second scale. This technology originated as way of utilizing superconductors for large-scale, ultrafast, digital, classical computing \cite{likharev2012superconductor}. At each time slice there either is a pulse or not, which implies that the unitary evolution is governed by two unitaries $\hat{U}_1$ and $\hat{U}_0$. Hence, the pulse train can be stored as a digital bit string with 0 and 1 denoting no pulse and a single pulse respectively. SFQ devices are interesting candidates for quantum computation since they potentially allow for ultrafast gate operations as well as scalable quantum hardware \cite{li2019scalable}. We model the pulses as $\Delta t = 2.0 \si{\pico \second}$ Gaussian functions $	
	 \Omega(t) =  \frac{a}{\sqrt{2\pi}\tau} e^{-\frac{(t-\Delta t/2)^2}{2\tau^2}},$ where $\tau = 0.25\si{\pico \second}$ and $a = 2\pi/1000$. The pulse is depicted to the right in Fig.~\ref{fig:SFQ}a. 
	 
	The optimization task is to find the input string that maximizes the fidelity functional \eqref{eq:Fidelity}. The current approach for this type optimization is to apply a genetic algorithm (GA) \cite{sutton1994genetic,whitley1994genetic,liebermann2016optimal}. Besides GA and AlphaZero, we also compare two conventional algorithms, Q-learning and stochastic descent (SD) as in Ref. \cite{bukov2018PhasesReinforcement}. Q-learning was one of the first RL algorithms developed, and applied recently to quantum control \cite{bukov2018PhasesReinforcement, bukov2018reinforcementFloquet}. It is a tabular-based algorithm that applies one-step updates in order to solve the optimal Bellman equation \cite{watkins1992q} (see Methods). SD is a time-local, greedy optimizer that changes the pulse at a randomly chosen time if this results in an increasing fidelity.

Our unified AlphaZero algorithm has an action space of 60 for the neural network, and thus we group together binary SFQ action choices of multiple time steps.  For this purpose, we take larger steps in time, and the 60 action choices are given using bit strings from a randomly chosen basis (see Methods). We benchmark the different algorithms by using equal wall-time simulations. For all simulations presented in this paper, we used a wall-time of 50 hours on an Intel Xeon X5650 CPU (2.7 GHz) processor. Similar to Ref. \cite{liebermann2016optimal} we use a population size of 70 with a mutation probability of 0.001 for the GA (see Methods). 
	
	The results are plotted in Fig.~\ref{fig:SFQ}b. Amongst conventional approaches, we see the SD algorithm performs slightly better than the GA. We attribute this to the fact that the SD algorithm is a greedy exploitation algorithm, while the GA is an exploration algorithm performing random permutations. As with many exploration algorithms, learning can be quite slow. Meanwhile, the Q-learning algorithm performs especially poorly. However, this algorithm is a tabular-based method. Such methods are known to break down for larger search spaces. This is one reason why modern RL algorithms use deep neural networks instead, motivating also our use of AlphaZero. We emphasize that AlphaZero also contains a deep lookahead tree search, which we found crucial to the success of our RL implementation (having also tested DQN \cite{mnih2015human} against simpler problems).  We see in Fig.~\ref{fig:SFQ}b that AlphaZero indeed performs dramatically better than the greedy approach, with over an order of magnitude improvement in the low error regime. We attribute this drop in error to the existence of a quantum speed limit (QSL) at or near 60ns, a minimum time for high-fidelity computation. This regime is known to be the most computationally difficult to optimize, with a high probability of local trapping \cite{Moore2012,bukov2018PhasesReinforcement,sorensen2018quantum}.

    AlphaZero and GA are both learning algorithms in the sense that they utilize previous obtained solutions in order to form new ones. We  compare the learning curves for the two algorithms in Fig.~\ref{fig:SFQ}c, where we have plotted the infidelity as a function of wall time at $60\si{\nano \second}$. For AlphaZero, we use the infidelity after each episode, where each data point is unique. For GA, we use the best performing solution in the population after each iteration. Since GA is a relatively greedy algorithm it performs very well initially, but fails to explore the larger solution space as the members in the population converge upon a single class of solution and the learning curve flattens out. In contrast, AlphaZero keeps a high level of exploration that ultimately allows it to reach a very large number of different high-fidelity solutions.  
	
	\subsection{Constrained analog pulses}

		\begin{figure}[t]
		\includegraphics[]{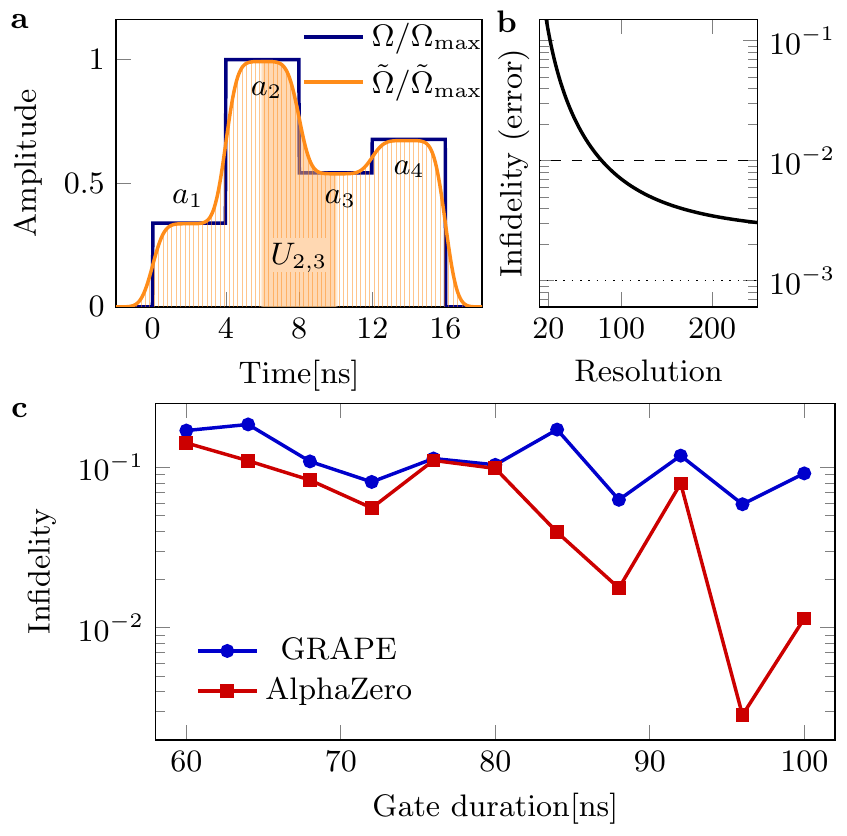}
		\caption{\textbf{a} A piecewise constant pulse (dark blue) convoluted by a Gaussian filter (light orange). Here $\sigma = 0.7 \si{\nano \second}$. \textbf{b} The error of the unitary as a function of its resolution. \textbf{c} Comparison between AlphaZero and GRAPE on the cross resonance gate using Gaussian filtered pulses.}
		\label{fig:pulse_with_error}
	\end{figure}
	
	A common challenge within quantum optimization is achieving realistic and efficient controls when experimental limitations constrain the underlying dynamics. Such constraints become very important when high precision is required, e.g.~for very high fidelity operation of quantum technologies. Here, we consider standard constraints on duration, bandwidth, and maximum energy. Such constraints can be expected to greatly increase the computational cost of Hessian approximation-based solutions, which are otherwise known to converge quickly \cite{ Rabitz2004Traps} and generally outperform other greedy methods \cite{machnes2011comparing, sorensen2018quantum}.  The workhorse algorithm for this is GRAPE \cite{khaneja2005optimal}, with quasi-Newton \cite{quantumLBFGS} and exact derivative \cite{MotzoiGRAPE} enhancements being crucial to the state of the art and its super-linear convergence.
	
	We model the bandwidth constraints via a convolution with a Gaussian filter function
	\begin{align}
	\tilde{\Omega}(t) =  \int_{-\infty}^{\infty} e^{-\frac{(t-t')^2}{\sigma^2}}
	\Omega(t') dt',
	\label{eq:GaussianConvolution}
	\end{align}
	where $\tilde{\Omega}(t)$ denotes the filtered control function. Fig.~\ref{fig:pulse_with_error}a illustrates the effect of this filter. Here, a piecewise constant pulse (dark blue) with amplitudes $a_{1-4}$ is convoluted into a smooth pulse (light orange) via Eq.~(\ref{eq:GaussianConvolution}). Throughout the remainder of this paper, we constrain the pulse amplitude to lie between $0$ and $\Omega_{\text{max}}/2\pi = 1.0 \si{\giga \hertz}$.

	Most commonly, GRAPE is applied to piecewise constant pulses, but it can be modified to include filtering \cite{MotzoiGRAPE,Kirchhoff2018CRG}, as we also do here. Each time-step is divided into a number of substeps (giving the resolution) and the filtered pulse is then approximated as being constant within each substep. This subdivision is depicted in Fig.~\ref{fig:pulse_with_error}a as light orange vertical lines. In order to obtain the gradient, GRAPE calculates the time-evolution unitary using matrix exponentiation at each substep. 

Fig.~\ref{fig:pulse_with_error}b shows the error (infidelity) between the exact and discretized unitaries as a function of the resolution. If we seek errors below the desired gate error ($10^{-2}$), the resolution should be around a couple of hundred. This significantly impedes the performance of GRAPE for this type of problem, since it requires considerably more matrix multiplications. 
A different strategy is to limit the control to a set of discretized amplitudes whose corresponding unitary can be calculated in advance and then apply a discretized optimization algorithm such as AlphaZero. In order to do so, we apply a two-action update strategy, where we propagate from half the previous pulse to halfway into the next one. So, if the previous action was $a_2$ and the next one $a_3$ then the unitary $U_{2,3}$ would correspond to the shaded region in Fig.~\ref{fig:pulse_with_error}a. Here we ignore negligible contributions from adjacent pulses. For instance, calculating $U_{2,3}$ would be independent of $a_1$ and $a_4$. Here we limit the amplitude to 60 different values (out of a continuous set), hence this methods requires calculating $60^2=3600$ unitaries, which we do in the beginning of the simulation.  

In our comparison between AlphaZero and GRAPE, we choose $ 4.0 \si{\nano \second}$ convoluted pulses using $\sigma = 0.7 \si{\nano \second}$. For GRAPE, we choose a resolution of 200. Fig.~\ref{fig:pulse_with_error}c shows the results of an equal wall-time simulation. Here, AlphaZero obtains a systematic improvement over its domain-specialized counterpart. At $96 \si{\nano \second}$, AlphaZero outperforms GRAPE with an improvement that is significantly above one order of magnitude. Interestingly, both graphs shows significant fluctuations, which we attribute to the difficulty of the optimization task itself caused by the highly constrained dynamics. This is likely compounded by the random initialization of the neural network which can effect the convergence properties of AlphaZero. Despite these fluctuations, AlphaZero performs significantly better in the regime of interest corresponding to infidelities below $10^{-2}$. 
\subsection{Piecewise-constant analog pulses}

	\begin{figure}[t]
		\centering
		\includegraphics[]{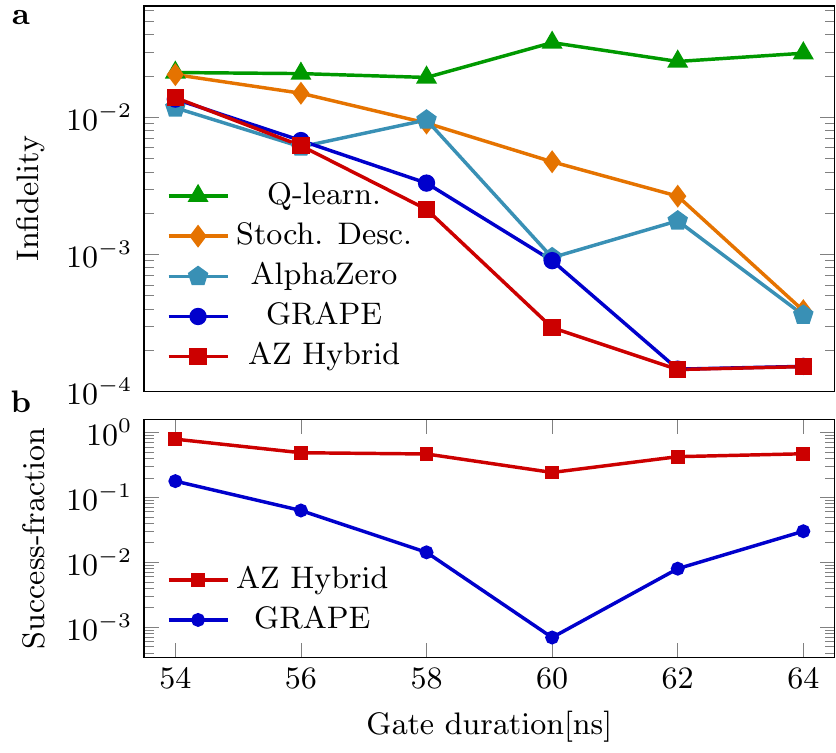}
		\caption{\textbf{a} An equal wall-time comparison between the various algorithms. The AlphaZero (here abbreviated AZ) Hybrid is presented in the text. \textbf{b} The fraction of successful solutions found by AlphaZero Hybrid and the GRAPE algorithm.}
		\label{fig:CRGResults}
	\end{figure}
	
	\begin{figure*}[t]
	\centering
	\includegraphics[]{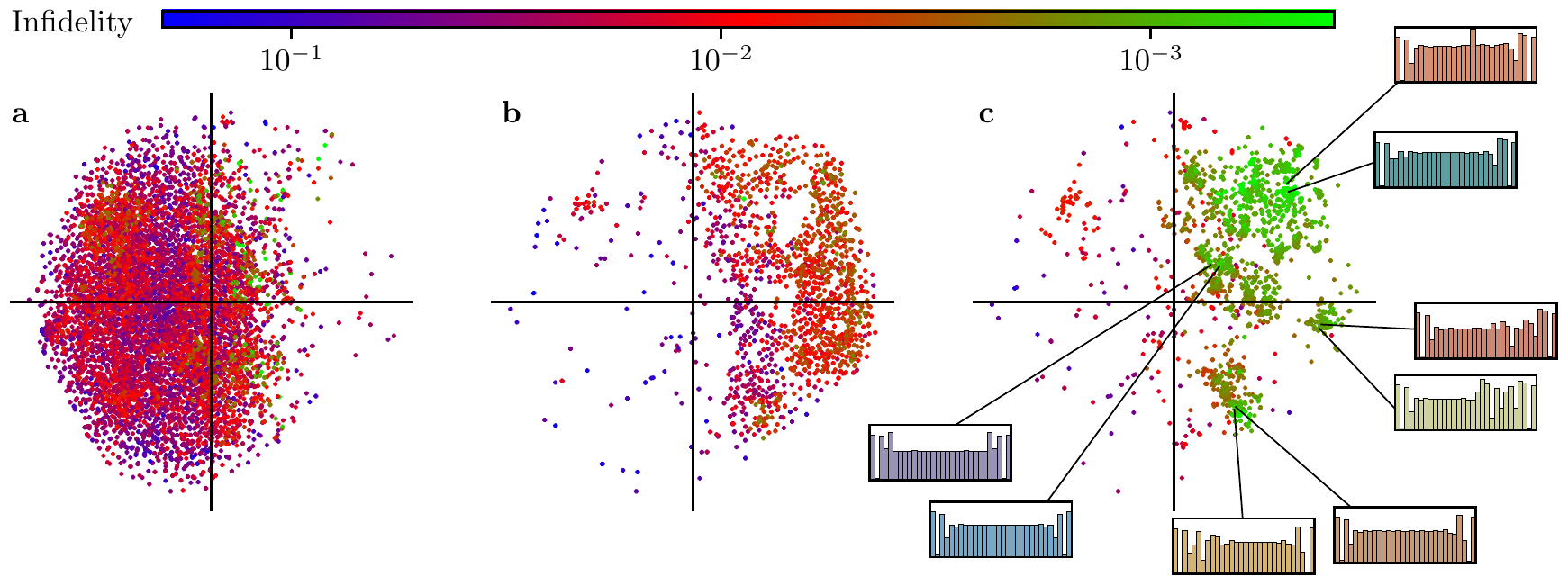}
	\caption{Two-dimensional representation of the final pulse vectors at $60\si{\nano \second}$ using the t-SNE algorithm. The color scale shows the infidelity of the pulses. \textbf{a} GRAPE with random seeding, \textbf{b} AlphaZero, \textbf{c}  The Hybrid, i.e.~AlphaZero solutions after being optimized with GRAPE. In the latter case, some example high-fidelity pulses are shown. }
	\label{fig:tSNE}
\end{figure*}
	
	So far, we have considered problems where gradient searches have not been applicable (digital sequence) or where gradient searches become inefficient (constrained analog pulses). For specific tasks where highly specialized algorithms exist and are known to perform relatively well, domain-agnostic algorithms typically perform inadequately. Thus, to properly benchmark our algorithm we have also considered the domain of piecewise constant pulses, a scenario where GRAPE typically performs extremely well due to the presence of high-frequency components and the limited number of matrix multiplications. In the following we hence focus on picewise constant pulses where we choose a single step duration of $2 \si{\nano \second}$. 
	
	In this scenario, we characterize the performance of the exploitation and exploration algorithms in terms of both the variety of solutions found and the quality of the solutions. At first, we compare the algorithms already discussed, namely Q-learning, Stochastic Descent, AlphaZero, and GRAPE. Fig. \ref{fig:CRGResults}a shows GRAPE is able to outperform the other algorithms for piecewise constant pulses. However, AlphaZero still performs well despite its limitation of only having amplitude-discretized controls. To improve the AlphaZero algorithm further we conceive a hybrid algorithm where GRAPE optimizes the solutions found by AlphaZero. The hybrid algorithm, which is given the same wall-time as the others, is also plotted in Fig.~\ref{fig:CRGResults}a. Here the hybrid algorithm shows a significant improvement over GRAPE near $60 \si{\nano \second}$, which we again relate to the presence of a quantum speed limit where the optimization task becomes difficult due to induced traps in the fidelity landscape. It is also worth noting that the optimization curve flattens out and the two algorithms again perform equally well when the pulse duration goes beyond $62 \si{\nano \second}$. We attribute this to the existence of a secondary QSL, i.e.~further improvement below $10^{-4}$ in infidelity requires gate durations beyond $200 \si{\nano \second}$ (not plotted here). 
	
	We also quantify the number of successful solutions found by either GRAPE or the hybrid AlphaZero algorithm, which we define as solutions having infidelities within four times the lowest infidelity obtained. The fraction of successful solutions are plotted in Fig.~\ref{fig:CRGResults}b. Here the improvement is even more substantial. At $60 \si{\nano \second}$, we find almost three orders of magnitude more successful solutions compared to GRAPE with random seeding. The fact that the GRAPE-curve dips around $60 \si{\nano \second}$ seems to confirm our previous statement about the QSL in the sense that this is a combinatorially harder region to obtain relatively good solutions. Having a large number of good solutions is especially important because experimentally it may be that some are better suited or some provide additional advantages.
	
	To further investigate the differences between the two algorithms, we compare the exploration of the control parameter landscape using a two-dimensional embedding provided by the t-SNE visualization method \cite{maaten2008visualizing, Sorensen2016}. 
	
	We do a single t-SNE analysis at $60 \si{\nano \second}$, plotted in Fig.~\ref{fig:tSNE}, which we have separated for clarity into different figures for GRAPE (a), the Hybrid before optimization (b), and after optimization (c). Here the color scale depicts the infidelity. Strikingly, the two algorithms seem to prefer entirely different portions of the landscape. GRAPE mostly finds solutions to the left in the t-SNE representation, but its high performing solutions are actually to the right. Interestingly, AlphaZero primarily finds solutions in the right region, which implies that AlphaZero has identified an underlying basic generic structure of good solutions. When all the AlphaZero solutions are optimized this leads to a large quantity of high performing solutions that inhabit the same region in the t-SNE representation. 
	
	We also see that the hybrid solutions naturally cluster towards some general basins of attraction. This suggests that AlphaZero has not converged on a single class but multiple different classes of solutions with different underlying physics. Some pulses from different clusters are depicted, showing some resemblance to typical bang-bang sequences. The different clustering that occurs demonstrates that a global exploration has indeed taken place, effectively finding different classes of solutions in different parts of the landscape.

\begin{figure}[h]
		\centering
		\includegraphics[]{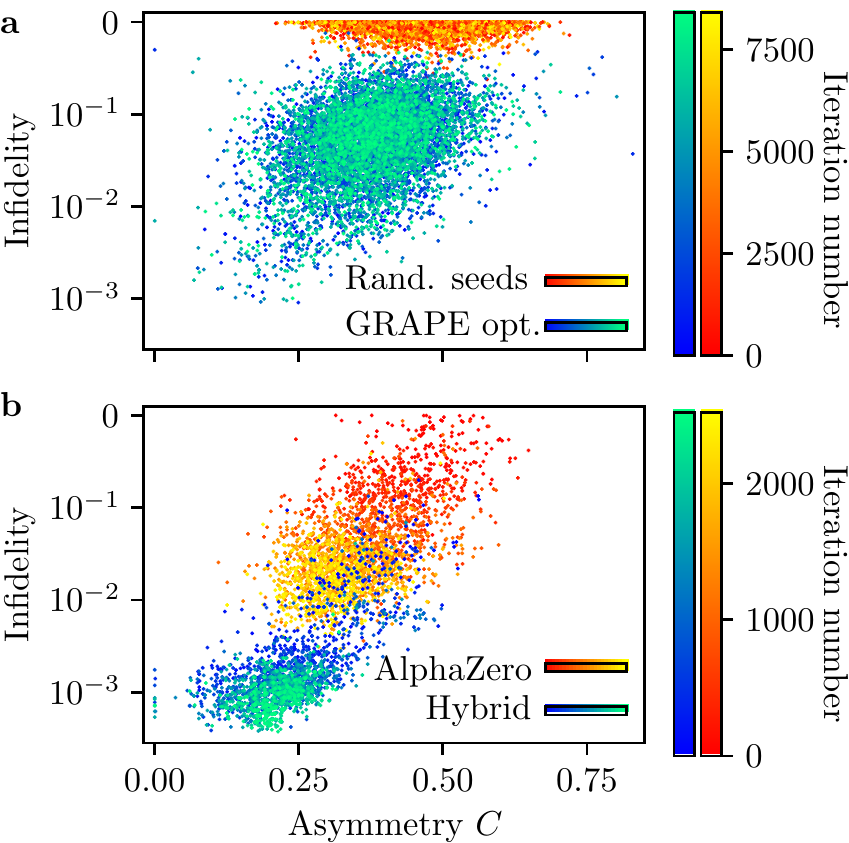}
		\caption{The initial seeds and the GRAPE optimization at $60\si{\nano \second}$ for \textbf{a} random generated seeds and \textbf{b} AlphaZero's solutions i.e. the Hybrid. The figures plot the infidelity ($1-\mathcal{F}$) as a function of the asymmetry measure \eqref{eq:Assymetry}. The color scale depicts the iteration of the algorithm.}
\label{fig:Symmetry}
\end{figure}

We further test the hypothesis that AlphaZero has found underlying structure that supersedes a shallow heuristic search. Note that the solutions seem to have at least some symmetry with respect to a reflection around the center of the time-axis. In fact, this symmetry already exists in the control problem. Since the Hamiltonian is real and the target its own transpose, the fidelity is unchanged if the pulse sequence is reversed i.e.~$\mathcal{F}(\Omega_1, \Omega_2, \ldots, \Omega_{N-1}, \Omega_N) = \mathcal{F}(\Omega_N, \Omega_{N-1}, \ldots, \Omega_2, \Omega_1)$. However, it is not \emph{a priori} clear that satisfying this symmetry is a good control strategy. We quantify the degree of time-asymmetry in the pulses via the measure
\begin{align} \label{eq:Assymetry}
C(\{\Omega(t)\}) = \frac{1}{N}\sqrt{\sum_{j=1}^{N} |\Omega_j-\Omega_{N-j}|^2},
\end{align}
where $C = 0$ implies pulses that are completely palindromic, i.e.~symmetric with respect to reversion of the sequence.  

We plot in Fig.~\ref{fig:Symmetry} the infidelity and the asymmetry for the two algorithms i.e. for GRAPE using random seeding (a) and the Hybrid, i.e.~GRAPE using the AlphaZero solutions (b). Here the color scale depicts the iteration number. The first thing to notice is that high fidelity solutions tend to maintain this symmetry. The second feature is that GRAPE often only partially satisfies this symmetry. In contrast, AlphaZero learns over its training to increasingly prefer this symmetry, moving towards the bottom left of the plot. After post-optimization using GRAPE, the solutions improve significantly in infidelity and move ever further to the bottom left emphasizing this trend.
We conclude that AlphaZero has identified this underlying symmetry specific to the problem instance we have chosen. Naturally, hard-coding such heuristics would not only be inefficient, but for many problems finding symmetries is nontrivial. Using deep learning, AlphaZero is able to learn these hidden symmetries without the need for human intervention. We therefore expect that AlphaZero's ability to learn hidden problem structures generalizes to other problems as well.
\section{Discussion}

From our three examples, we conclude that the AlphaZero methodology of combining neural network and guided tree search reinforces global information about good solutions that can also mark a significant algorithmic advantage for quantum optimization. This is true for specific problems, but especially when comparing across a range of problems. None of the other algorithms we have considered are able to do well on all three problems, be it with heuristic, machine learning or domain-specialized approaches.

The three problems considered marked different optimization tasks, but AlphaZero is able to find high fidelity solutions with a single set of algorithmic hyperparameters.  This suggests that learning the control landscape can be performed with minimal expert knowledge about the physical problem.

This conclusion is further enforced by the realization that hidden symmetries in the dynamics can be effectively learned by AlphaZero during its training. Such unexpected symmetries are not trivial to find for many Hamiltonians and would require significant human intervention even where they can be found.  More over, hard-coding such heuristics into optimization algorithms can have many pitfalls, limiting broad exploration and potentially leading to suboptimal trapping in the optimization landscape.

Nonetheless, because the deep exploration methodology is by design agnostic to expert knowledge, it is most powerful when combined with specialized knowledge about locally exploiting promising seeds, leveraging the vast body of literature about local quantum optimization.  This tradeoff between exploitation and exploration is a common trend in reinforcement learning and optimization in general. For example, in AlphaZero's chess matches with its competing AI, Stockfish \cite{Stockfish}, the latter was trained with sophisticated domain knowledge and thus was generally acknowledged as outperforming in the final moves of games. Combining the domain-agnostic exploration of the former with the domain-specialized exploitation of the latter seems like a common sense solution, as we have done here in the quantum dynamics case.  An even tighter integration of the two approaches that examines the tradeoffs during different learning stages may also be promising.  Alternatively, one could also also relax the \emph{tabula rasa} character of the learning to enhance the exploration abilities using specialized knowledge. Supervised learning can in principle speed up the initial learning phase, perhaps most seemlessly when integrated with other broad exploration strategies, for instance crowd sourcing \cite{heck2018remote, Sorensen2016}.

In this work we have considered digital, constrained, and underconstrained optimization of controlled quantum dynamics in the context of the design and execution of physical quantum-mechanical devices.  This choice was deliberately made because the most advanced algorithms exist in this field owing to half a century of dedicated research.  That being said, many of the more abstract and potentially groundbreaking dynamics algorithms, including those used in the design of digital sequences of quantum circuits or for analog evolutions in annealers and variational eigensolvers, can be seen as direct analogues of the algorithmic framework illustrated here.

\section{Methods}
\subsection*{Reinforcement Learning}
A general RL setup consists of an environment and an agent. At each time step $t$, the environment is characterized by a state $s_t$. Given $s_t$, the agent selects an action $a_t$ that changes the environment to a new state $s_{t+1}$. Based on this change the agent receives a feedback signal called a reward, $r_{t+1} \in \mathbb{R}$. The agent must learn how to maximize the sum of rewards it receives during an episode. This is done by implementing a policy $\pi$, which is a mapping from all states of the environment to probabilities of selecting possible actions $\text{Pr}(a|s)=p_a(s)$. The state-value function describes the quality of a given policy
\begin{align}
v_{\pi}(s) = \mathbb{E}_{\pi} \left[\sum_{t'>t} r_{t'} \Bigg| s=s_{t'} \right], 
\label{eq:value_function}
\end{align}
which is simply the expected sum of future reward staring from state $s$ and subsequently following the policy $\pi$. Given two policies $\pi$ and $\pi'$ we say that $\pi \geq \pi'$ if $v_{\pi}(s) \geq v_{\pi'}(s)$ for all states $s$. 

The task considered here is to a construct a pulse sequence, which realizes a target unitary. At each time step, the agent must select an action that updates the unitary representing the state of the system. At each time step, the reward is zero except at the last step where it is simply the fidelity given by equation (\ref{eq:Fidelity}).

\subsection*{AlphaZero implementation \label{sec:AlphaZero} } 
AlphaZero is a policy improvement algorithm that combines a neural network with a Monte Carlo Tree Search (MCTS) as depicted in Fig.~\ref{fig:Transmon_treesearch_neuralnetwork} b and c \cite{silver2017mastering,silver2017masteringchess}. The neural network maps from states to policies $\mathbf{p} = (p_1, p_2, \ldots)$ and values $v$. The MCTS, guided by the neural network, also computes a policy $\boldsymbol{\pi}$ that the actions are drawn from. At each time step, the policy $\boldsymbol{\pi}$ is stored in a replay buffer. At the end of an episode, the final score $z = \sum_t r_t$ is also stored in the buffer. Training of the neural network uses data drawn uniformly at random from the replay buffer in order to let the network predictions $(\mathbf{p}, v)$ approach the stored data $(\boldsymbol{\pi}, z)$. This is done by minimizing the loss function

\begin{align}
l(\boldsymbol{\theta}) = (z-v)^2 - \boldsymbol{\pi}^T \log \mathbf{p} +c||\boldsymbol{\theta}||^2,
\end{align}
where the last term denotes L2 regularization with respect to the network parameters $\boldsymbol{\theta}$. 

A MCTS is a way of looking several steps ahead by only visiting a small subset of possible future states. The tree is built by nodes (states) connected to each other by edges (state-action pairs). Each edge has four numbers associated with it: The number of visits $N(s,a)$, the total action value $W(s,a)$, the mean action value $Q(s,a)$, and a prior probability of selecting set edge $P(s,a)$. Starting from the root node (initial state), a single tree search moves through the tree by selecting actions according to $a_t = \arg \max_a (Q(s_t,a)+U(s_t,a))$, where $U(s_t,a)$ denotes an uncertainty given by
\begin{align}
U(s,a) = c_{\text{puct}}P(s,a)\frac{\sqrt{\sum_b N(s,b)}}{N(s,a)}.
\end{align}
Here $c_{\text{puct}}$ denotes a parameter determining the level of exploration. If a terminal node or a leaf node (i.e. a not-previously-visited state) is encountered, the search stops. The tree is expanded in the latter case by adding the node and initializing its edges as $N(s,a)=W(s,a)=Q(s,a)=0$ and $P(s,a)=p_a$, where $p_a$ is given by the neural network. The rest of the tree is updated by using the state-value $v$ in a backwards pass through all the visited edges since the root node according to $N(s,a) \leftarrow N(s,a) + 1$, $W(s,a) \leftarrow W(s,a) + v$, and $Q(s,a) \leftarrow W(s,a)/N(s,a)$. After a pre-set number of such searches have been conducted, an actual policy is calculated according to

\begin{align}
\pi(a|s_0) = \frac{N(s_0,a)^{1/\tau}}{\sum_b N(s_0,b)^{1/\tau}},
\end{align}
where $s_0$ is the root state and $\tau$ denotes a parameter controlling the level of exploration, which is annealed during the simulations. The action in drawn from the policy and the rest of the tree is reused for subsequent searches during the episode. 

For all tasks presented in this paper we used the same algorithmic parameters. The learning rate was $0.01$,  $c_{\text{puct}} = 1.0$, and $\tau$ was hyperbolically annealed from $1.0$ using an annealing rate of $0.001$. After $\tau$ was annealed below a value of $0.90$ we switched to deterministic policies by setting the largest policy value to one and the others zero. The neural network was a simple feed forward network where the hidden nodes consisted of four layers. Each layer contained $400$ nodes followed by batch normalization and a rectified linear unit. Both the policy and the value head of the neural network consisted of a single hidden layer as well, where the policy head ended in a sigmoid-layer with same dimension as the action space and the value head ended in a single linear node. The L2 regularization parameter was $c = 0.001$ and we used stochastic gradient descent (SGD) for training the network. Similar to the AlphaZero paper \cite{silver2017mastering} we achieve more exploration by adding Dirichlet noise to the search probabilities for the root nodes $P(s,a) = (1-\epsilon)p_a + \epsilon\eta_a$, where $\boldsymbol{\eta} \sim \text{Dir}(0.03)$ and $\epsilon = 0.25$.  

\subsection*{GA implementation} \label{sec:GA}
A genetic algorithm (GA) works by iteratively updating a population of solutions, which are bit strings \cite{sutton1994genetic,whitley1994genetic}. A GA generates new solutions based on the old population via processes inspired by biological evolution, namely crossover and mutations, which respectively combine two parent solutions and flip individual bits at random. If any improved solutions are found, these replace the worst ones in the population. Similar to Ref. \cite{liebermann2016optimal}, we used a population size of 70 and a mutation probability of 0.001. At each iteration we would select $2\times 30$ parent solutions.

\subsection*{Q-learning implementation} \label{sec:Qlearning}
Similar to equation (\ref{eq:value_function}) one can define an action-value function

\begin{align}
    q_{\pi}(s,a) = \mathbb{E}_{\pi} \left[\sum_{t'>t} r_{t'} \Bigg| s = s_{t'}, a = a_{t'} \right],
\end{align}
which is the expected reward if we choose action $a$ from state $s$ and then follow the policy $\pi$ \cite{sutton2011reinforcement}. Q-learning is a tabular-based RL algorithm, which approximates the optimal action-value function i.e. the action-value function for the optimal policy $\pi_{opt} = \max_{\pi} v_{\pi}(s)$. The approximation $Q(s,a)$ is initialized at random and subsequently updated according to

\begin{align}\nonumber
    Q(s_t,a_t) &\leftarrow Q(s_t,a_t)\\ &+ \alpha [r_{t+1} + \max_{a_{t+1}} Q(s_{t+1},a_{t+1}) - Q(s_t,a_t)],
\end{align}
where $\alpha$ denotes the learning rate. Similar to Ref. \cite{bukov2018PhasesReinforcement} we choose our state to be a tuple of time and control $s = (t,\Omega)$. The learning rate was $\alpha = 0.001$ and we followed an epsilon-greedy strategy with linear annealing of epsilon \cite{sutton2011reinforcement}. 

\subsection*{Cross Resonance Gate}
The cross resonance (CR) gate \cite{paraoanu2006microwave,chow2011simple, magesan2018effective} is currently the standard fixed-frequency qubit entangling gate used on transmon systems.  Its main advantage is avoiding the overhead associated with magnetic (flux) tuning of the frequency \cite{Groszkowski2011,koch2007charge}, which can be a leading cause of dephasing. As  illustrated in Fig. \ref{fig:Transmon_treesearch_neuralnetwork}a, the physical setup we optimize includes two fixed frequency qubits that are coupled to each other via a transmission line resonator. The transmons \cite{koch2007charge} may be modelled as anharmonically spaced Duffing oscillators \cite{Khani2009nonlin}, resulting in an extended Jaynes-Cummings model Hamiltonian 
\begin{align}\nonumber 
\label{eq:drift_Hamiltonian}
H &= \sum_{j=1,2}\bigg( \omega_j \hat{b}_j^{\dagger} \hat{b}_j
+ \frac{\delta_j}{2} \hat{b}_j^{\dagger}\hat{b}_j(\hat{b}_j^{\dagger}\hat{b}_j-1)
+ \omega_r \hat{a}^{\dagger}\hat{a}
\\ &+  g_j (\hat{b}_j^{\dagger} \hat{a} 
+ \hat{b}_j\hat{a}^{\dagger}) + \Omega(t)(\hat{b}_j+\hat{b}_j^{\dagger}) \bigg), 
\end{align}
where $\hat{b}_{1,2}^{\dagger}(\hat{b}_{1,2})$ and $\hat{a}^{\dagger} (\hat{a})$ are the transmon and cavity creation (annihilation) operators respectively. Here $\omega_1 \neq \omega_2$ is the transmon resonance frequency, $\delta_{1,2}$ denotes the anharmonicity, $\omega_r$ denotes the cavity resonance, and $g_{1,2}$ the transmon-cavity coupling. The transmons are directly driven by external control parameters $\Omega(t)$, increasing the controllability compared to earlier architectures that drive through the common cavity. The transition of the second qubit is then driven resonantly through the control line of the first \cite{Kirchhoff2018CRG}. 

This model may be significantly simplified using the method in Ref. \cite{magesan2018effective}.  After adiabatic elimination of the cavity and block diagonalization into the qubit subspace, the authors derive an equivalent equation (Eq.~3.3), which is the same as our Eq.~(\ref{eq:Hamiltonian}).

To see that the natural gate that is produced from this Hamiltonian is a $\sqrt{ZX}$ gate, a (Schrieffer-Wolff) perturbative expansion shows \cite{chow2011simple} that the leading coefficients in the effective driving terms are given by
\begin{equation}
    H_d = \Omega(t) \big[ XI + \frac{J}{\Delta} ZX + m IX\big],
\end{equation}
where $Z$ and $X$ are Pauli matrices acting on the respective qubits, $I$ is the identity, and $m$ is a hand-tuned crosstalk parameter. The single-qubit terms and higher order terms (not shown) must be decoupled in the control optimization in order to correctly implement the CR gate.

\subsection*{Digital pulses} \label{sec:Methods_digital_pulses}

For each time step, the evolution of the system is governed by either one of two unitaries $\hat{U}_0$ and $\hat{U}_1$, which respectively corresponds to the amplitude being zero or not \cite{leonard2019digital}. We calculate these unitaries in advance by solving the Schrödinger equation numerically. The entire pulse sequence can be encoded as a bit string as illustrated to the left in Fig.~\ref{fig:SFQ}a and the corresponding unitary can be calculated as $\hat{U}(T) = \prod_{j = 1}^{N} \hat{U}_{b_j}$ where $b_j \in [0,1]$. Pulse durations in the nano-second scale require $10^4-10^5 $ steps.

For AlphaZero we create 60 unitaries $\{ \hat{U}^{(i)} \}_{i=1}^{60}$ by drawing a bit string $b_1^{(i)},b_2^{(i)},\dots, b_{500}^{(i)}$ at random, where $  b_j^{(i)}\in [0,1]$, which we then multiply $\hat{U}^{(i)} = \prod_{j=1}^{500} \hat{U}_{b_j^{(i)}}$. In order to obtain pulse sequences that have both high and low concentrations of $b_j^{(i)} = 0$ we anneal the probability $\text{Pr}(b_j^{(i)} = 0)$ linearly from one ($i = 1$)  to zero ($i = 60$). The 60 unitaries constitute the action space and the unitary is now calculated as $\hat{U}(t) = \prod_{t' \leq t} \hat{U}^{(a_{t'})}$. The 60 actions allows us to use the same neural network architecture as for piece-wise constant and filtered pulses which have the same input space dimension.

\bibliography{refs}

\section*{Acknowledgements}
This work was funded by the European Research Council, John Templeton Foundation, and the Carlsberg Foundation. The authors would also like to personally thank Jesper H. M.~Jensen, Carrie Ann Weidner, and Miroslav Gajdacz for fruitfull discussions and input. The numerical results presented in this work were obtained at the Centre for Scientific Computing, Aarhus http://phys.au.dk/forskning/cscaa/.

\section*{Author contributions}
M.D. and F.M. wrote and implemented the software, performed the simulations, and analyzed the data. All authors contributed to interpreting data as well as participating in useful scientific discussions. J.S. planned and supervised this project. J.J.S. advised the project on a day-to-day basis. All authors contributed to writing this paper.

\section*{Code and data availability}
All data presented in this paper and the code that generated it is available upon request. All requests should be directed at J. Sherson (sherson@phys.au.dk). 
\end{document}